\documentclass[structabstract]{aa}  

\usepackage[varg]{txfonts}
\usepackage{longtable,lscape}
\usepackage{graphicx}
\usepackage{natbib}
\bibpunct{(}{)}{;}{a}{}{,} 

\newcommand {\Msun} {\, \mbox{M$_{\odot}$}}
\newcommand {\AU} {\, \mathrm{AU}}
\newcommand {\COri} {\mbox{$\theta^1\mathrm{C}\:\mathrm{Ori}$}}

\begin{document}
 
\title{Disc-mass distribution in star-disc encounters}

\author{M. Steinhausen
          \inst{1,2},
           C. Olczak \inst{3} \inst{,4,} \inst{5}
        \and
         S. Pfalzner
           \inst{1}}

          \institute{\inst{1} Max-Planck-Institut f\"ur Radioastronomie (MPIfR), Auf dem H\"ugel 69, 53121 Bonn, Germany\\
            \inst{2} I. Physikalisches Institut, Universit\"at zu K\"oln, Z\"ulpicher Str. 77, 50937 K\"oln, Germany\\
           \inst{3} Astronomisches Rechen-Institut (ARI), Zentrum f\"ur Astronomie der Universit\"at Heidelberg (ZAH) M\"onchhofstr. 12-14,
           69120 Heidelberg, Germany \\
            \inst{4} Max-Planck-Institut f\"ur Astronomie (MPIA), K\"onigstuhl 17, 69117 Heidelberg, Germany\\
           \inst{5} National Astronomical Observatories of China (NAOC), Chinese Academy of Sciences (CAS), 20A Datun Lu, Chaoyang District, Beijing 100012, China\\
            \email{mstein@mpifr-bonn.mpg.de}
          }

   \date{Received      ; accepted     }

 
  \abstract
  {}
  {Investigations of stellar encounters in cluster environments have demonstrated their potential influence on the mass and angular momentum of protoplanetary
    discs around young stars. In this study it is investigated in how far the initial surface density in the disc surrounding a young star influences the outcome of an encounter.}
   {The numerical method applied here allows to determine the mass and angular momentum losses in an encounter for
     {\emph{any}} initial disc-mass distribution. Based on a power-law ansatz for the surface density, $\Sigma (r) \propto r^{-p}$, a parameter study of
     star-disc encounters with different initial disc-mass distributions has been performed using N-body simulations.}
   {It is demonstrated  that the shape of the disc-mass distribution has a significant impact on the  quantity of  the  disc-mass and angular momentum losses in star-disc encounters.  Most sensitive are the results where the outer parts of the disc are perturbed by
  high-mass stars.   By contrast, disc-penetrating encounters lead more or less independently of the disc-mass
    distribution always to large losses. However, maximum losses are generally obtained for initially flat distributed
  disc material. Based on the parameter study a fit formula is derived, describing the relative mass and angular momentum loss
  dependent on the initial disc-mass distribution index $p$.  Generally encounters lead to a steepening of the
    density profile of the disc. The resulting profiles can have a r$^{-2}$-dependence or even steeper independent of the initial
    distribution of the disc material.}
{From observations the initial density distribution in discs remains unconstrained, so the here demonstrated strong
   dependence on the initial density distribution might require a revision of the effect
   of encounters in young stellar clusters. The steep surface density distributions induced by some encounters might be the prerequisite to form planetary systems similar to our own solar system.}
   \keywords{Methods: numerical -- Protoplanetary discs -- (Stars:) circumstellar matter}
   \maketitle


\section{Introduction}
\label{sec:introduction}

There is increasing observational evidence that most, if not all, stars are initially surrounded by a circumstellar disc. For
example, \citet{2000AJ....120.3162L} found by examining the L-band excess of young stars in the Trapezium cluster a fraction of $80-85\%$ of all stars to be surrounded by
discs. With time  these protoplanetary discs  become depleted of gas and dust and eventually
disappear \citep{2001ApJ...553L.153H, 2002astro.ph.10520H, 2006ApJ...638..897S, 2007ApJ...671.1784H,
  2008ApJ...672..558C, 2008ApJ...686.1195H, 2009AIPC.1158....3M, 2010A&A...516A..52M}. Currently it is unclear which of a variety of physical mechanisms dominates this evolutionary processes,
among them there are internal processes like viscous
torques \citep[e.g.][]{1987ARA&A..25...23S}, turbulent effects \citep{2003ApJ...582..869K} and magnetic fields
\citep{2002ApJ...573..749B}, while photoevaporation \citep{2001MNRAS.325..449S,
  2001MNRAS.328..485C, 2003ApJ...582..893M, 2004RMxAC..22...38J, 2005MNRAS.358..283A, 2006MNRAS.369..229A, 2008ApJ...688..398E,
  2009ApJ...699L..35D, 2009ApJ...690.1539G} and tidal torques \citep{1993ApJ...408..337H, 1993MNRAS.261..190C, 1994ApJ...424..292O,
  1995ApJ...455..252H, 1996MNRAS.278..303H, 1997MNRAS.287..148H, 1997MNRAS.290..490L, 1998MNRAS.300.1189B, 2004ApJ...602..356P,
  2005ApJ...629..526P, 2006ApJ...653..437M, 2008A&A...487..671K} are candidates for external disc destruction processes.

The focus of the present numerical investigation is on the effect of gravitational star-disc interactions on the disc-mass distribution
and, therefore, the mass and angular momentum of the discs. A star-disc encounter\footnote{The term 'star-disc encounter' means here encounters in which only one of
  the stars is surrounded by a disc, in contrast to disc-disc encounters, which denominate encounters in which both stars are
  surrounded by discs.}  can cause matter to  become unbound, be captured by the perturbing star or pushed inwards and potentially be
accreted by the central star. The extent to which this happens depends on the periastron distance, the mass ratio of the stars,
the eccentricity and, moreover, the initial (pre-encounter) mass distribution in the disc. 
So far, limitations in observations of protoplanetary discs  prevented to  reveal consistently how the surface density of low-mass discs
develops. This means that due to observational constraints there does not exist a unique predetermined initial state for the disc-mass distribution before an
encounter. A wide variety of surface densities have been derived from fitting resolved millimeter continuum or
line emission data with parametric disc structure models \citep[e.g.][]{1996ApJ...464L.169M, 1997ApJ...489..917L} or in combination with broadband
spectral energy distributions (SED) \citep{2000ApJ...534L.101W, 2001ApJ...554.1087T, 2002ApJ...566.1124A, 2002ApJ...581..357K, 2007ApJ...659..705A}. While those
studies have profoundly shaped our knowledge of disc structures, they are all fundamentally limited by low angular resolution.

The standard fitting method underlies the assumption that the surface density $\Sigma$ has a simple power-law dependence of the form 

\begin{equation}
  \Sigma(r) \propto r^{-p}
  \label{eq:surface_density_p}
\end{equation}
out to some cut-off radius
\citep[e.g.][]{2007ApJ...671.1800A}. Estimates based on numerical power-law models fitted to
observational data lead to distribution indices $p$ ranging roughly from $p = 0$ to $p = 2$. Recent works
even found anomalous results, e.g.  
\citet{2009ApJ...701..260I} observed distribution indices of $p < 0$.

Analytical approaches suggest different disc-mass distribution indices, too. The most widely used model is that for a steady state viscous accretion disc with
a surface density distribution index of $p = 1$ \citep[e.g.][]{1998ApJ...495..385H}. However, simulating
the evolution of protostellar discs that form self-consistently from the collapse of a molecular cloud core lead to a surface distribution
index of $p = 1.5$ \citep{1990ApJ...358..515L, 2005A&A...442..703H, 2007MNRAS.381.1009v} while treating magnetized disc material results in a flatter
disc-mass distribution of $p = 0.75$ \citep{2007ApJ...665..535S}. 
Taking this huge variety of distributions into account, one has to consider different initial disc-mass distributions to evaluate their effect in star-disc encounters.

Most previous numerical studies of star-disc encounters used only a single density distribution, mainly focusing on
the case of a theoretically motivated $r^{-1}$ disc-mass distribution \citep{1996MNRAS.278..303H, 1997MNRAS.287..148H,
  2004ApJ...602..356P, 2006ApJ...642.1140O, 2006ApJ...653..437M, 2007A&A...462..193P}. Star-disc encounters with different initial
disc-mass distributions were only considered in a very limited way. \citet{1995ApJ...455..252H} performed numerical simulations of
two different mass distributions ($p = 0$ and $p = 1.5$) concentrating on parabolic encounters with equal mass  stars  while
\citet{1997MNRAS.287..148H} investigated initial surface distributions of $p = 0$ and $p = 1$ for close and penetrating encounters
of an unity mass ratio. A study of a wide parameter range  focusing  on multiple initial disc-mass distributions is still lacking.

Nevertheless, numerical studies of star-disc encounters only allow a sectional view on the processes since it is not possible to simulate each
combination of encounter parameters. Earlier analytical studies by \citet{1994ApJ...424..292O} did not suffer from this
shortcoming. In her study a first order approximation of the angular momentum loss dependent on the initial disc-mass distribution
is given. However, the validity of her results is limited to large periastron radii (for example $r_{\mathrm{peri}} /
r_{\mathrm{disc}} > 3 $ for $M_2 / M_1 = 1$), where the angular momentum loss is usually well below $10 \, \% $. Close or even penetrating
encounters cannot be interpreted by  this linear perturbation theory \citep{1994ApJ...424..292O, 2005A&A...437..967P} making
numerical studies indispensable in this regime.

In this work the effects of  star-disc encounters have been investigated for a large parameter space  considering most configurations that can be expected in a
typical young cluster. The investigated mass distributions cover the entire range of the so far observed disc-mass distributions. 

Section $2$ shortly describes the numerical methods and parameter range used in this study while Section $3$ presents the results
of the simulations including a fit formula for the mass and angular momentum loss depending on the initial disc-mass distribution
index $p$. A summary and discussion is given in Section $4$.


\section{Methods}

 The encounter between a disc-surrounded star with a secondary star is modeled using a code that  is based on the numerical method described in \citet{2003ApJ...592..986P}.  In our simulations only one star is initially surrounded by a disc.  However, previous investigations showed that star-disc encounters
can be generalized to disc-disc encounters as long as there is no significant mass exchange between the discs \citep{2005ApJ...629..526P}. In case of a mass exchange the discs can be to some extent replenished, so that for very close encounters the mass loss determined in this study would be slightly overestimated. 

 A summary of the disc properties can be found in Table 1. 
The disc is represented by $10.000$ pseudo-particles, distributed according to a given particle distribution $\propto r^{-b}$. 
Choosing this relatively low number of simulation particles is motivated by the aim to cover a large encounter parameter space. However, performing test simulations with $50.000$ particles, shows that the lower resolution actually suffice for the here investigated properties. 

The simulation particles move initially on Keplerian orbits around the central star. The disc
extends from an inner gap of $10 \AU$ to $100 \AU$. The inner cut-off avoids additional complex calculations of direct star-disc interactions and saves computation
time. Any pseudo-particle that reaches a sphere of $1 \AU$ around the central star is removed from the simulation and stated in a
commonly used simplified approach as being accreted \citep{2002MNRAS.336..705B, 2005ApJ...633L.137V, 2008A&A...487L..45P}. The
vertical density distribution in the disc follows 

\begin{equation}
  \rho (r,z) = \rho_{0} \exp \left( - \frac{z^2}{2H^2(r)} \right),
\label{eq:vertical_density}
\end{equation}
where $\rho_{0}$ is the unperturbed mid-plane particle density on the equatorial plane with $\rho_{0} \propto r^{-(b+1)}$ and $H(r)$ the vertical half-thickness of the disc
\citep[see][]{1981ARA&A..19..137P}.  $H(r)$ is chosen  as $H(r) = 0.05 \, r$  according to a temperature profile of
  $T = T_0 \cdot r^{-1}$ with $T = 20 \mathrm{K}$ at the inner edge of the disc.

\begin{table}
  \caption[]{Initial conditions for all simulations}
  \label{tab:initialconditions}
  $$ 
  \begin{array}{p{0.5\linewidth}l}
    \hline
    \noalign{\smallskip}
    variable  & \mathrm{value} \\
    \noalign{\smallskip}
    \hline
    \noalign{\smallskip}
    Outer disc radius,  $r_{\mathrm{disc}}$  & 100 \, \mathrm{AU}  \\
    Inner disc radius, $r_{\mathrm{i,disc}}$  &  10 \AU  \\
    Number of particles,  $n_{\mathrm{part}}$  & 10.000 \, (50.000) \\
    Disc-mass,  $m_{\mathrm{disc}}$  & 10^{-4} \Msun \, (10^{-3} \Msun) \\
    Particle distribution index, $b$ & 0 \mathrm{, } \;  ( \frac{7}{4} ) \\
    Mass distribution index, $p$ & 0 \mathrm{, } \; \frac{1}{2} \mathrm{, } \; 1 \mathrm{, } \; \frac{7}{4} \\
    \noalign{\smallskip}
    \hline
    \noalign{\smallskip}
 Rel. perturber mass, $M_{2} / M_{1}$ & 0.1 - 500 \\
    Rel. periastron distance, $r_{\mathrm{peri}} / r_{\mathrm{disc}}$   & 0.1  - 20 \\
    Eccentricity,  $e$  & 1 \\
    Simulation time, $t_{\mathrm{end}}$ & \sim 3000 \, \mathrm{yrs}  \\
    \noalign{\smallskip}
    \hline
  \end{array}
  $$ 
\end{table}

The temporal development during the encounter is modeled using  a fifth-order Runge-Kutta Cash-Karp integrator
with an adaptive time step size control for the numerical calculations. Long-range interactions of the gravitational forces between the
disc particles and the perturbing star are calculated using a hierarchical tree method
\citep{1986Natur.324..446B}.

The total simulation time of $t_{\mathrm{end}} \approx 3000$ yrs for each encounter corresponds to three orbital periods of the
outermost particles in the investigated standard disc with size $r_{\mathrm{disc}} = 100 \AU$ around a $1 \Msun$ star. This
time span was found to be adequate for the calculation of all relevant quantities.  

The discs obtained at the end of the simulations are still developing in the sense that taking into account viscous forces the
eccentric orbits of perturbed disc pseudo-particles would re-circularize on time scales probably in excess of $10^5 \mathrm{yr}$. However, bound and unbound material already can be clearly distinguished shortly after the encounter.

 In this study we are interested in protoplanetary discs, which are usually of low mass compared to protostellar
  discs \citep{2011MNRAS.tmp.1310B}. Here low-mass discs mean $m_{\mathrm{disc}} \ll 0.1 M_{1}$, where $M_{1}$ is the mass of the disc surrounded
  star. The actual used disc-mass value is $10^{-4} M_{1}$. However, we tested as well the case $m_{\mathrm{disc}} = 10^{-3} M_{1}$
  and found no differences in the result when including self-gravity in the simulations. In addition, we performed test
  simulations including pressure and viscous forces using a SPH code (for a description of the code see
  \citet{2003ApJ...592..986P}), but only found negligible differences ($<3\%$) in the results. Therefore, in most of our simulations self-gravitation and viscous forces were neglected in favor of higher computational performance \citep{2003ApJ...592..986P, 2004ApJ...602..356P}.

The actual values of mass and angular momentum loss induced by the encounters are obtained by averaging over several simulations
with different seeds for the initial particle distribution. The errors in the mass and angular momentum losses are determined as
the maximum differences between the simulation results. They typically lie in the range of $2 - 3 \, \%$.

Besides these obvious performance benefits the possibility to neglect the self-gravitation and pressure and viscous forces allows
to further optimize computation time by treating the disc particles as pseudo-particles without a fixed mass during the simulation.

 The standard method to generate the initial disc-mass distribution in the simulations is by assigning each pseudo-particle the same mass
\citep{1998MNRAS.300.1189B, 2004ApJ...602..356P, 2006ApJ...642.1140O, 2007A&A...462..193P}. This approach requires to perform the
whole suite of simulations afresh for each variation of the initial disc-mass distribution.  By contrast, here the different disc-mass distributions are realized by  using a fixed pseudo-particle distribution and  assigning their masses \emph{after} the simulation process according to the desired density distribution in the disc. The implementation of such  a flexible numerical scheme allows using one suite of simulations for any initial disc-mass distribution.
Note, that the post-processing of the particle mass is only possible because of the restriction to low-mass discs where
self-gravitation and viscous forces can be neglected.

The particle masses are assigned to each pseudo-particle in the diagnostic step to calculate the encounter-induced losses and final mass distributions dependent on the initial distribution. 
For the example of a power-law disc-mass distribution with index $p$ the mass of a particle, $m_{\mathrm{i}}$, depends on its
radial position in the disc $r_{\mathrm{i}}$, the total disc-mass $m_{\mathrm{disc}}$, the number of pseudo-particles
$n_{\mathrm{part}}$, their radial positions $r_j$ and the underlying power-law particle distribution with
index $b$. Since the number of pseudo-particles is limited, the discrete particle masses $m_i$ are obtained by  

\begin{equation}
m_{\mathrm{i}} = \left ( \sum_{j=1}^{n_{part}} r_j^{-p+b} \right )^{-1} \cdot m_{\mathrm{disc}} \cdot r_{\mathrm{i}}^{-p+b} \mathrm{.}
\label{eq:mass_distribution}
\end{equation}
The so established independence of mass and particle distribution allows to improve the resolution of our simulations by placing most of the pseudo-particles initially where the interaction between the stars and disc is strongest.

In most of our simulations a constant particle distribution ($b = 0$) is used. This is done because the effects on mass and angular momentum are largest at the outskirts of the disc.
Using a constant pseudo-particle density ($b = 0$) throughout the disc, ensures a higher resolution in the outer parts even for
steep mass density profiles compared to the standard approach. By contrast, if one would be interested in processes that mainly
concern the inner parts of the disc, like for example accretion, it would be preferential to use not a constant particle
distribution but one that guarantees a high resolution close to the star. 
 
We have tested the method of a 
posteriori mapping of the particle masses against  representative test calculations for different initial pseudo-particle
distributions and against the standard method. The differences in the results where in each case negligible and thus justify the
generalization of our results for a constant particle distribution ($b = 0$).

Apart from the density distribution the outcome of an encounter depends on the relative periastron distance $r_{\mathrm{peri}} / r_{\mathrm{disc}}$ and the
mass ratio between the involved stars $M_2 / M_1$, where $M_2$ is the perturber mass. 
The here investigated parameter space  - for a summary see Table 1 -  is chosen in such a way that it spans the entire range of encounters likely to occur in a typical young cluster, like the Orion
Nebula Cluster. Recent investigations show that the ONC with its high central
density might be typical for young clusters with a mass $> 1000 \Msun$ \citep{2009A&A...498L..37P}.
The lower limit of the perturber mass ratio was chosen to be $M_{2}/ M_1 = 0.1$ as for smaller mass ratios the influence of the
perturber becomes insignificant. The upper limit, $M_{2}/ M_1 = 500$, is determined by
the maximum possible mass ratio in the ONC, which is given by the hydrogen burning limit as lowest mass ($0.08 \Msun$) and $40 \Msun$, the
approximate mass of the $\COri$ system, which contains the most massive star in the ONC. 
The limit where perturbations at large distances become negligible depends on the perturber mass. The inner edge of the disc marks
the lower value of $r_{\mathrm{peri}} / r_{\mathrm{disc}} = 0.1$.


\section{Results}


Although in principle any mass distribution of circumstellar discs can be studied, exemplary four disc-mass distributions are investigated here
in more detail. To cover the entire range of numerically and observationally determined disc-mass distributions a constant mass
distribution ($p = 0$), representing the lower boundary of expected distributions, and a $p = 7/4$ distribution, providing an upper
limit, are considered. Additionally, a $p=1$ distribution is investigated for comparison to previous star-disc encounter results and a $p = 1/2$ distribution,
which is in the range of analytical results for low-mass discs \citep{1973A&A....24..337S, 1981ARA&A..19..137P} and similar to
results found in recent investigations considering magnetized material (see Sec.~\ref{sec:introduction}).

The present investigation focuses on coplanar, prograde encounters on parabolic orbits ($e=1$)  (see example in
Fig.~\ref{fig:encounter}$a$) . Previous studies have shown that even for clusters as dense as the inner ONC region, most encounters in star clusters are expected to be close to parabolic \citep{1990ASSL..162..461L, 1994ApJ...424..292O,
  2010A&A...509A..63O}. Such parabolic encounters provide the strongest impact of the perturber on the disc since for higher
eccentricities the perturber only interacts shortly with the star-disc system and is, therefore, unable to influence the disc
significantly. The limitation to a certain orientation is more restricting, as inclined and retrograde encounters can lead to
lower losses in mass \citep{1993ApJ...408..337H, 1996MNRAS.278..303H, 1993MNRAS.261..190C, 2005A&A...437..967P}. However,
\citet{2005A&A...437..967P} showed that up to an inclination of $45^{\circ}$ the mass loss of the disc is only slightly reduced in
comparison to a coplanar orbit. This means that our results have to be regarded as an upper limit for encounters at a different
inclination.

\subsection{Surface density distribution}
\label{subsec:Density distribution}

\begin{figure}[thbp]
  \centering
  \includegraphics[width=0.5\textwidth]{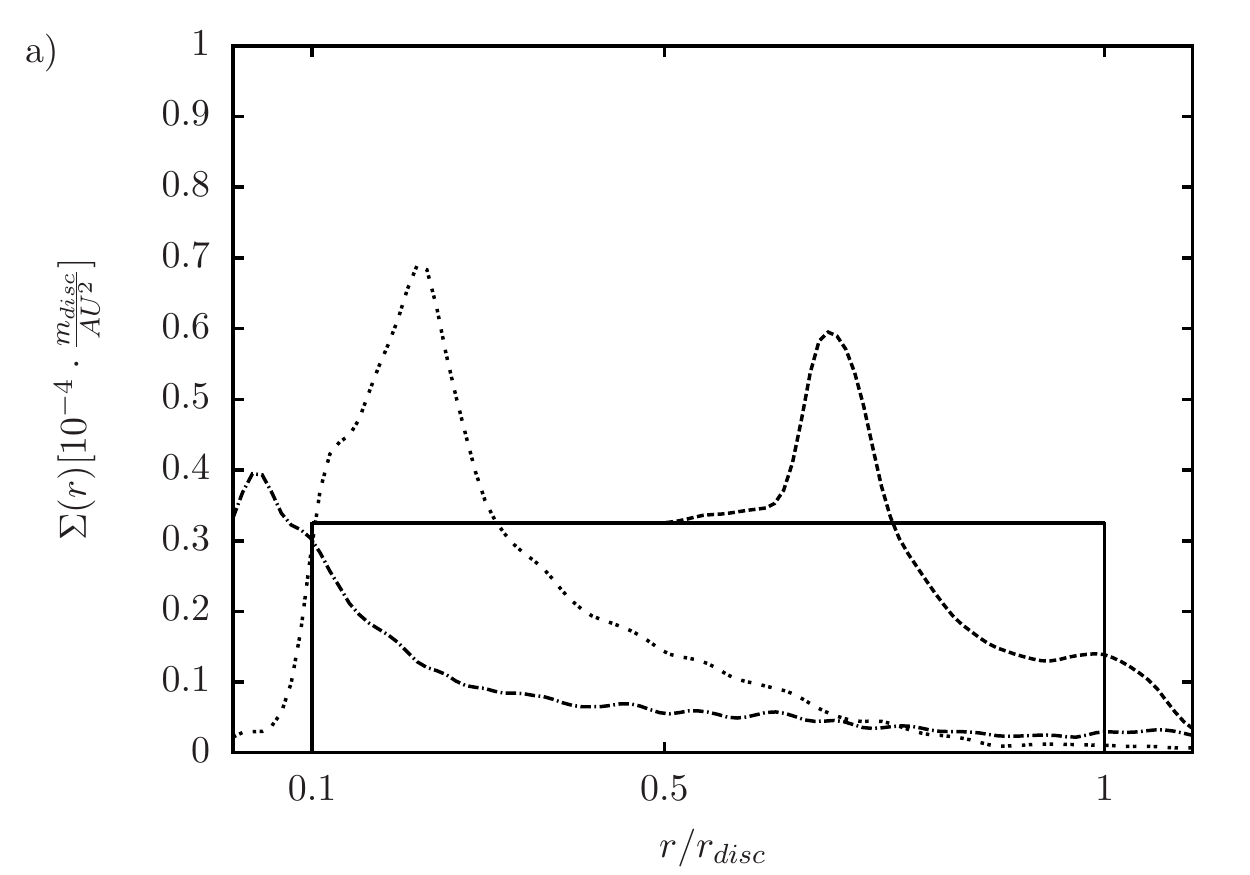}
  \includegraphics[width=0.5\textwidth]{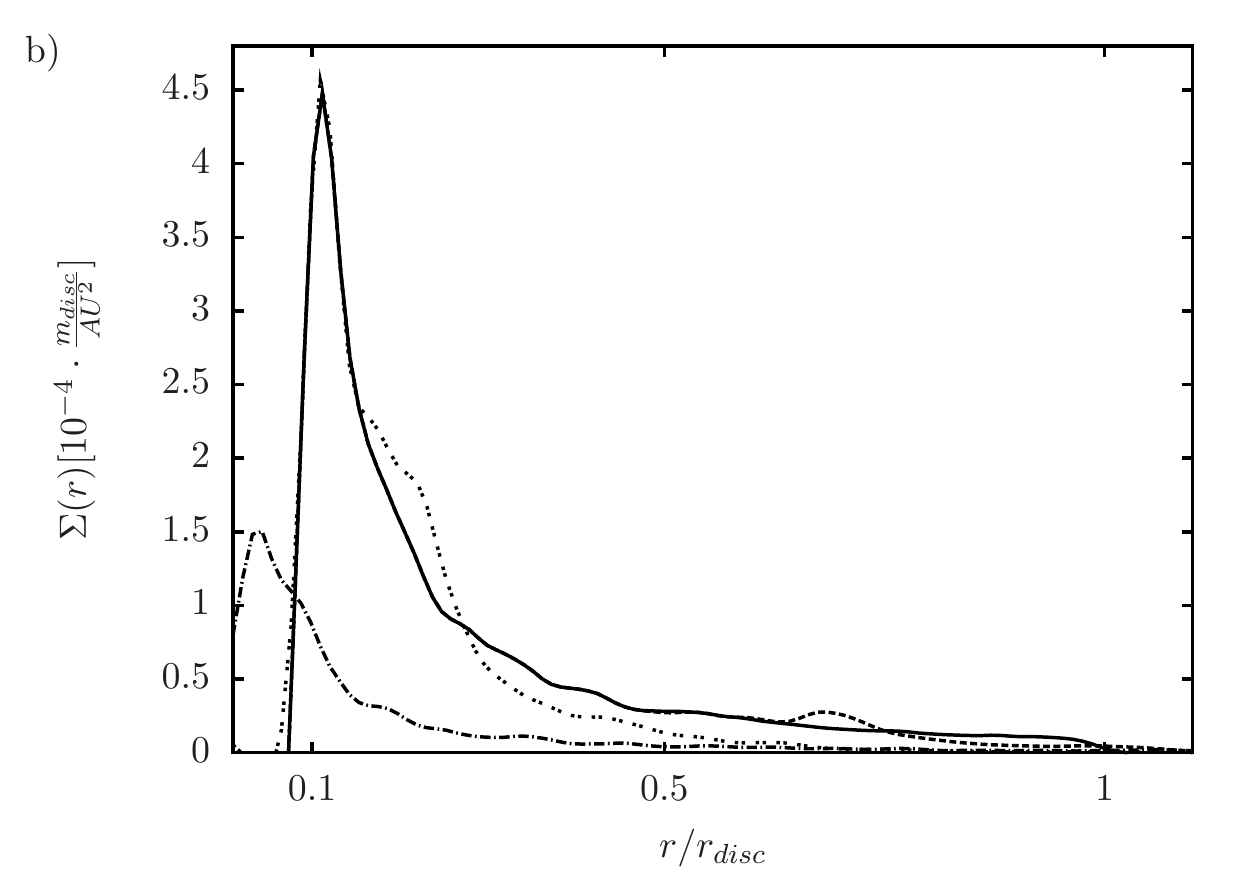}
  \caption{Initial (solid line) and final surface densities in case of $a)$ initially constant distributed disc material of $p = 0$ and $b)$ a steep distribution of $p
    = 7/4$. In both plots non-penetrating ($r_{\mathrm{peri}} / r_{\mathrm{disc}} = 3$, dashed line), grazing
    ($r_{\mathrm{peri}} / r_{\mathrm{disc}} = 1$, double-dotted line) and penetrating encounters ($r_{\mathrm{peri}} / r_{\mathrm{disc}} = 0.1 $, dashed-dotted line)
    are plotted for a perturbing star of equal mass ($M_2 / M_1= 1$).}
  \label{fig:surface_density}
 \end{figure}

To have a better understanding of the re-distribution of the disc material during a star-disc encounter in dependence of the initial disc-mass
distribution we first focus on the evolution of the surface density profiles.
The evolution of the surface density in star-disc encounters in dependence of the periastron distance for a
constant initial disc-mass distribution ($p = 0$) and one with a steep initial distribution ($p = 7/4$) can be seen in Fig.~\ref{fig:surface_density}. It shows the mass
distributions before (solid line) and after a penetrating ($r_{\mathrm{peri}} / r_{\mathrm{disc}} \leq 1 $, dashed-dotted line), grazing
($r_{\mathrm{peri}} / r_{\mathrm{disc}} = 1 $, double-dotted line) and distant encounter ($r_{\mathrm{peri}} / r_{\mathrm{disc}} = 3 $, dashed line). Here, the relative
perturber mass was chosen to be $M_2 / M_1 = 1$.

As known from previous investigations 
\citep{1997MNRAS.287..148H, 1997MNRAS.290..490L}  an encounter reduces the density in the outer parts of the disc by transporting  some part of the outer disc material inwards and some part migrating outwards. The latter might become captured by the
perturber and unbound if the encounter is strong enough.  These effects become more pronounced for small $r_{\mathrm{peri}} / r_{\mathrm{disc}}$ and/or  large $M_2 / M_1$. 
However, the actual amounts depend on the initial disc-mass distribution. 
In an equal-mass  distant  encounter ($r_{\mathrm{peri}} / r_{\mathrm{disc}} = 3$) most of the perturbed disc material is pushed inside
the disc towards lower disc radii. In this case $12.5 \, \%$ of the total disc-mass migrates inwards for an initial constant disc-mass
distribution ($p = 0$), whereas it is only $6 \, \%$ for an initial $p = 7/4$.

The effect of different initial disc-mass distributions becomes even more obvious for closer encounters like grazing
(double-dotted lines in Fig.~\ref{fig:surface_density}) or penetrating fly-bys
(dashed-dotted  lines in Fig.~\ref{fig:surface_density}). Here, the
migration process and also the differences between the initial distributions are dominated by material moving
outwards. In case of a grazing encounter and an initially constant disc-mass distribution $\sim 60 \, \%$ of the disc-mass
  becomes unbound. However, in addition $\sim 10 \, \%$ of the total disc-mass can be found outside the initial disc radius of
  $100$ AU still bound to the central star. By contrast, for the steep $p = 7/4$-mass distributions it only becomes $\sim 30 \,
  \%$ of the disc-mass unbound but again $\sim 10 \, \%$ is bound but situated outside $100$ AU. We conclude that generally the
  outer disc material is separated from the disc for grazing encounters while material initially located inside the disc ($r /
  r_{disc} \leq 0.7$) will be re-distributed but remains bound to the central star. Hence, prominent
  differences in encounter-induced disc losses for the investigated disc-mass distributions are expected for strong perturbations
  of the outer disc parts.
 In penetrating encounters part of the disc material is pushed further inside the disc resulting in an increased
surface density in the inner disc regions.  In extreme cases the disc loss can be increased to $ > 90 \, \%$ so that the final disc structure can no longer be regarded as a disc. 

\begin{figure}[htbp]
  \centering
  \includegraphics[width=0.5\textwidth]{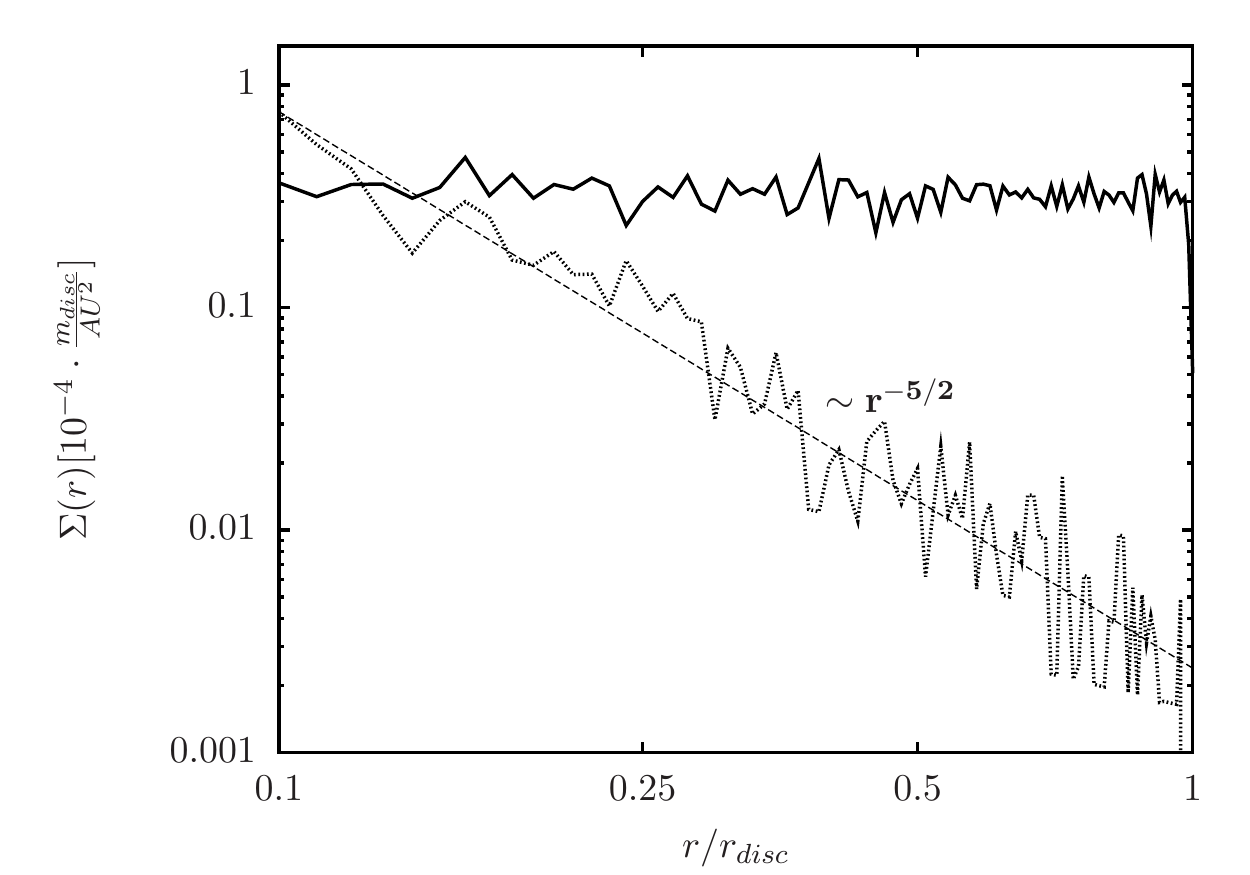}
  \caption{Initial ($p = 0$, solid line) and final surface density after an encounter of $M_2 /
    M_1 = 5.0$ and $r_{\mathrm{peri}} / r_{\mathrm{disc}} = 0.7$ (dotted line). The dashed straight line represents a slope of $p = 5/2$.}
  \label{fig:surface_log}
\end{figure}

In nearly all cases a steepening of the density profile is the general effect of an encounter. The degree of steepening depends on the mass ratio,
periastron distance and the initial disc-mass distribution. Fig.~\ref{fig:surface_log} shows that even initially flat
distributed disc-material ($p = 0$) can be redistributed into a surface density profile steeper than $p \approx 2$.

\begin{figure}[thbp]
  \centering
  \includegraphics[width=0.5\textwidth]{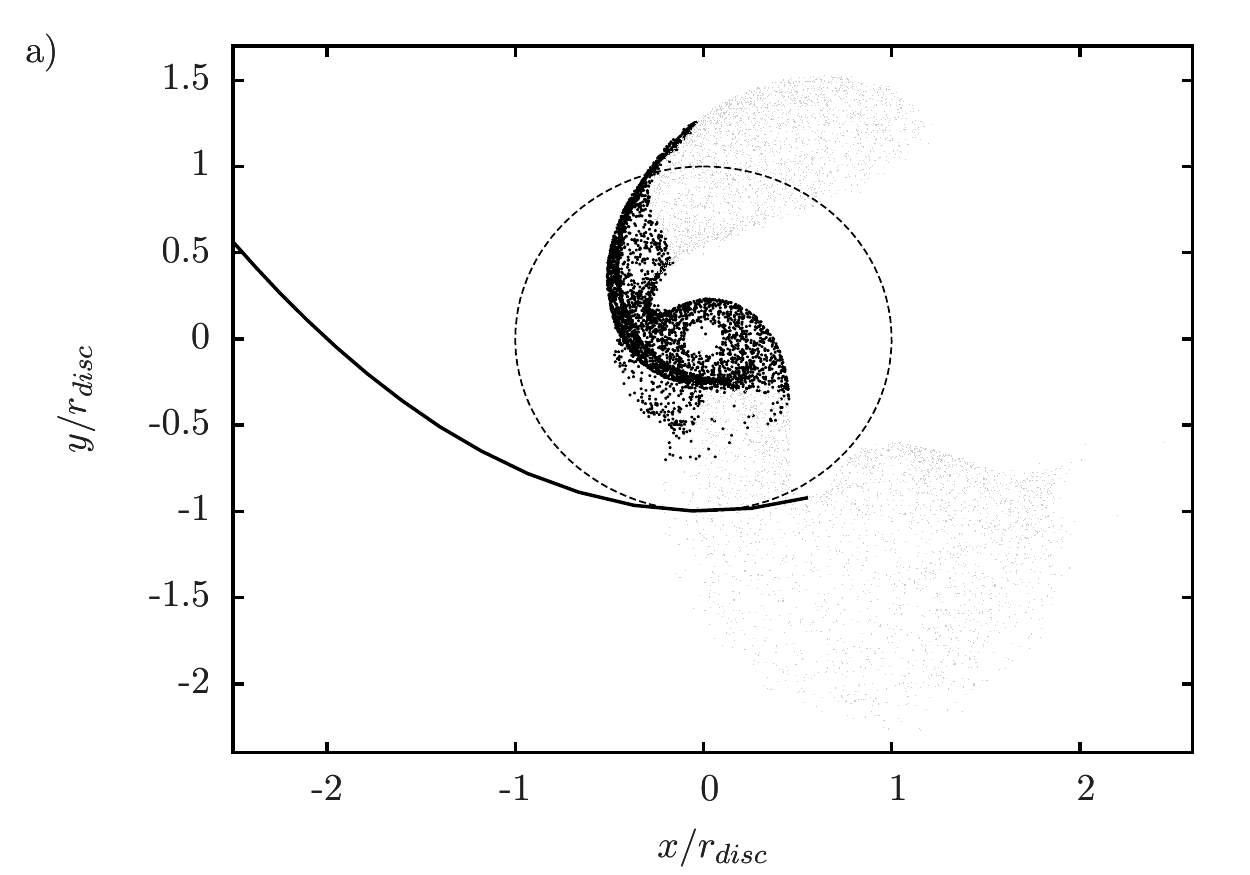}
  \includegraphics[width=0.5\textwidth]{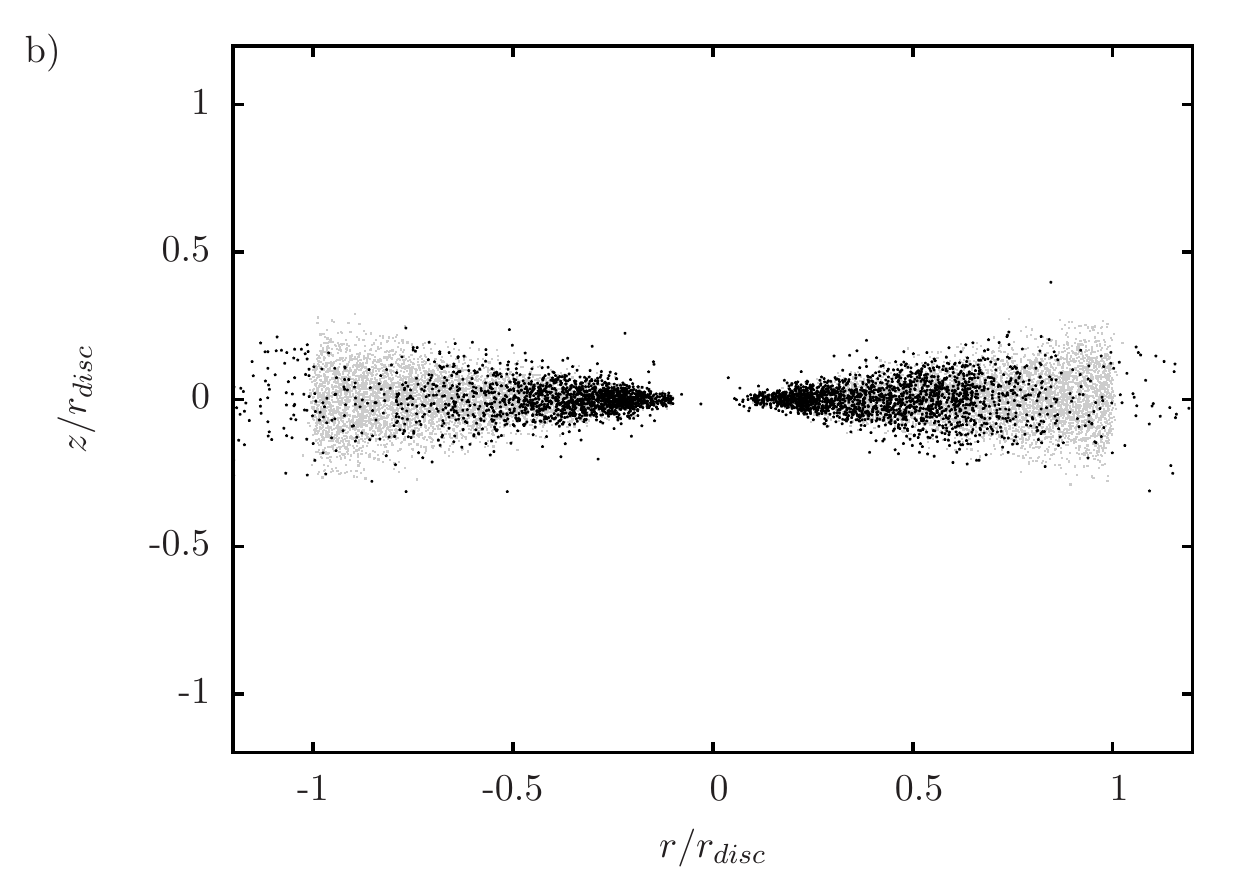}
  \caption{The dashed line in a) shows the boundaries of the initial disc while the solid line indicates the trajectory of a grazing
    perturber ($r_{\mathrm{peri}} / r_{\mathrm{disc}} = 1$) of equal mass ($M_2 / M_1 = 1$). Material that resides within the disc
    after the perturbation is marked as black squares while material that is in the end bound to the perturbing star, unbound or
    accreted is shown as gray dots. Note that the simulations were performed in three dimensions as can be seen in b).} 
  \label{fig:encounter}
\end{figure}

How is the vertical structure of the disc influenced by an encounter? As only coplanar encounters are investigated here, the
answer is: surprisingly little. Fig.~\ref{fig:encounter}$b$ shows the vertical particle profile of a strongly perturbed disc after a penetrating encounter of $M_2 /
  M_1 = 1$. The main effect of the encounter is a decrease in the number of particles in the outer disc regions. In these parts of the disc the resolution is somewhat diminished, but due
to our constant particle distribution ($b = 0$) still significantly larger than in previous investigations using $r^{-1}$ particle
distributions in their simulations.  

\begin{figure}[htbp]
   \centering
   \includegraphics[width=0.5\textwidth]{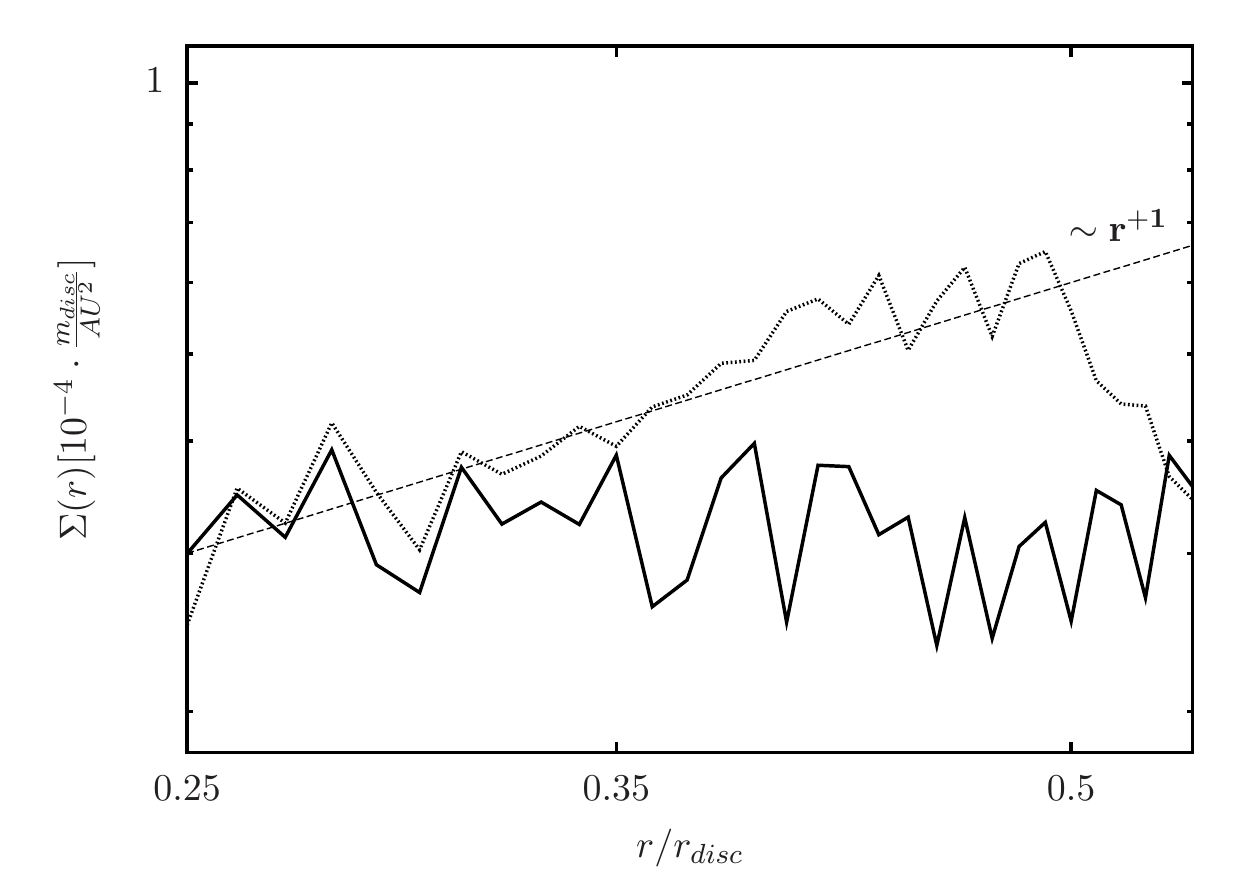}
   \caption{Initial ($p = 0$, solid line) and final surface density after an encounter of $M_2 /
     M_1 = 1.0$ and $r_{\mathrm{peri}} /  r_{\mathrm{disc}} = 2.0$ (dotted line). The dashed straight line represents a slope of $p = -1$.}
   \label{fig:surface_isella}
 \end{figure}

In a few special cases where material is being pushed moderately inwards even partially negative distribution indices as low as $p = -1$ can be the end product of an encounter (Fig.~\ref{fig:surface_isella}). The effect is most prominent for the inner part
of perturbed discs with initially constant distributed disc material. Nonetheless, the outer parts of the disc always represent a distribution with positive index, $p > 0$.
 
  This could perhaps explain the negative indices observed for some discs. Observations are still limited to measuring only part
  of the radial extension of the entire disc. If this happens to be the range were negative indices prevail, one would wrongly
  extrapolate negative indices for the whole disc, whereas in fact the overall index of the disc would still be positive.

\subsection{Relative disc-mass and angular momentum loss}
\label{subsec:Mass loss}


\subsubsection{Relative mass loss}
\label{subsec:Mass loss}

\begin{figure}[thbp]
\centering
\includegraphics[width=0.5\textwidth]{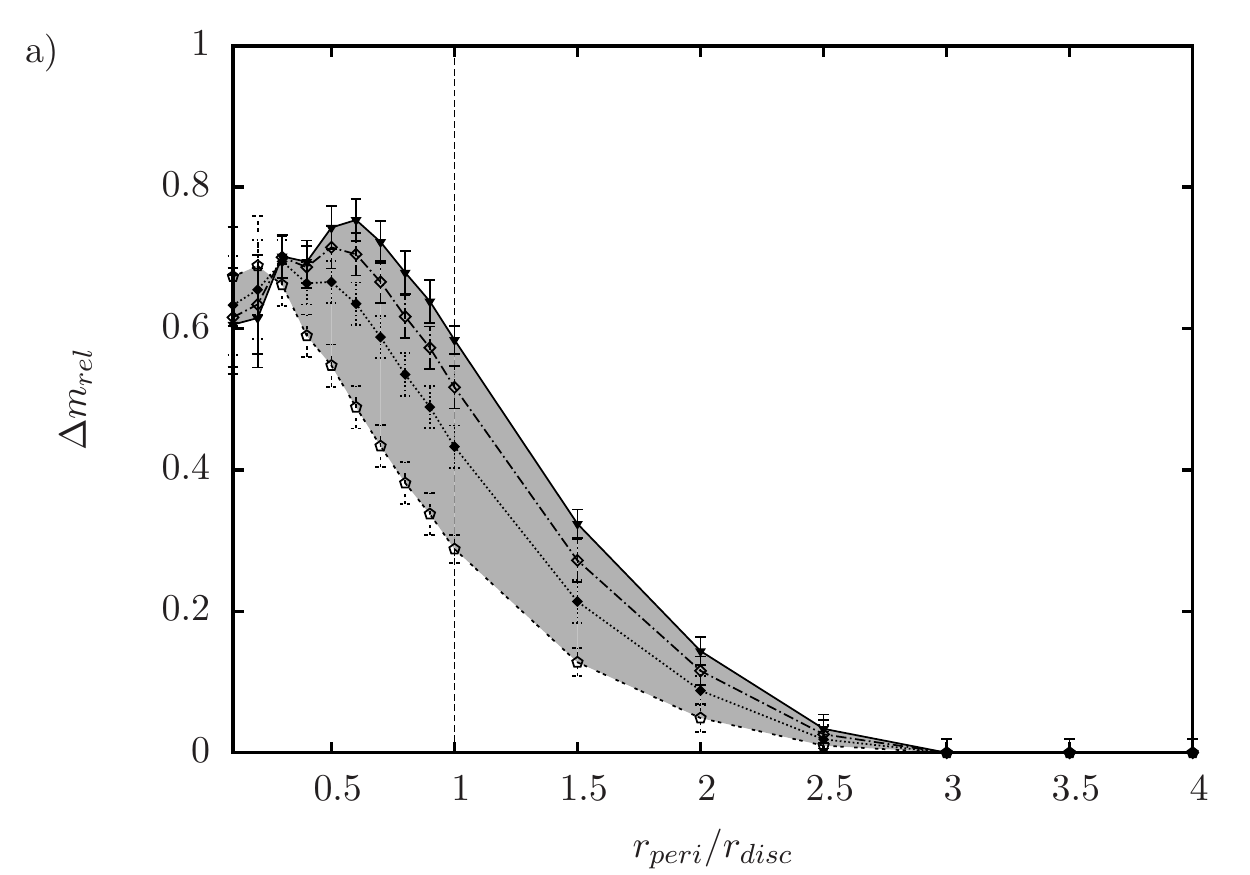}
\includegraphics[width=0.5\textwidth]{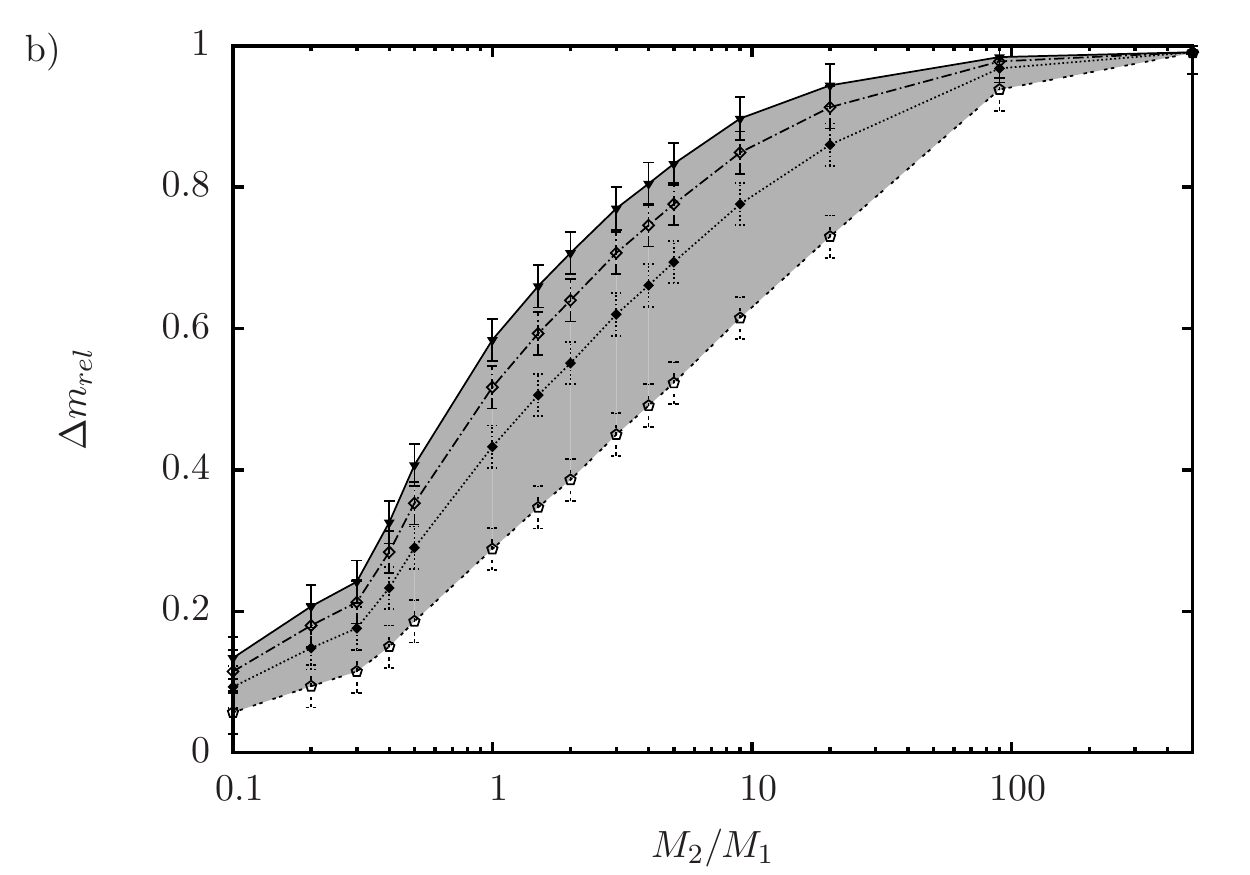}
\caption{The relative disc-mass loss of a $p = 0$ (solid line), $p = 1/2$ (dashed-dotted line), $p = 1$ (dotted line) and $p =
  7/4$ (double-dotted line) disc-mass distribution including all particles bound tighter to the central star than to the
  perturber and excluding unbound and accreted particles. The data is plotted for $a)$ different periastron distances and an
  equal-mass perturber and for $b)$ different perturber mass ratios and $r_{\mathrm{peri}} / r_{\mathrm{disc}} = 1 $. The
  vertical dashed line indicates the initial outer disc radius.}
\label{fig:angularloss}
\end{figure}

Besides changing the shape of the disc-mass distributions, star-disc encounters can remove material from the disc. Fig.~\ref{fig:angularloss} shows the relative disc-mass loss
$\Delta m_{\mathrm{rel}} = ( m_{t=0} - m_{t_{\mathrm{end}}}) / m_{t=0}$
for the four initial disc-mass distribution indices $p = 0$, $p = 1/2$, $p = 1$ and $p = 7/4$.
Two representative cases of orbital parameters are singled out: Fig.~\ref{fig:angularloss}$a$ shows the
dependence on the periastron distance for an equal-mass encounter, whereas Fig.~\ref{fig:angularloss}$b$ depicts the dependence on the
mass ratio for grazing encounters ($r_{\mathrm{peri}} / r_{\mathrm{disc}} = 1 $).

Qualitatively, the dependency of the relative disc-mass loss on the periastron distance follows previous results \citep{2005A&A...437..967P, 2006ApJ...642.1140O}. 
However, the absolute values change considerably for the different initial disc-mass distributions. The effect is largest for
nearly grazing encounters where mainly the outer disc regions are affected.  As to be expected from the respective fraction of material in the
outer regions,  maximum mass losses are obtained for initially constant disc-mass
distributions while losses for the $r^{-7/4}$-distribution are lowest. The largest difference in mass loss between
  the investigated disc-mass distributions in Fig.~\ref{fig:angularloss}$a$ occurs
for $r_{\mathrm{peri}} / r_{\mathrm{disc}} = 0.9 $, whereas the
 $r^{-7/4}$-distribution has only a mass loss of $33 \, \%$, the constant mass distribution has a mass loss of $64 \, \%$.

The situation is somewhat different for very  close penetrating encounters ($r_{\mathrm{peri}} /  r_{\mathrm{disc}} \leq 0.3$) where the
discs are so strongly perturbed that the resulting structure can hardly be described as a disc. In this
case the disc-mass loss seems relatively independent of the initial mass distribution (see Fig.~\ref{fig:angularloss}$a$).
At the other end of the parameter space - i.e. at large relative periastron distances - the mass loss becomes too small ($\Delta
m_{\mathrm{rel}}  \leq 10 \% $) to infer any dependence on the initial distribution.

 Similarly, the dependence on the initial mass distribution is less
pronounced for  $M_2 /  M_1 < 0.3 $ and $M_2 / M_1 > 90 $ (see Fig.~\ref{fig:angularloss}$b$ for a grazing encounter). So
generally weak perturbations - whether distant or of low mass ratio $M_2 / M_1$ - are not capable to significantly influence the discs, while
in the case of strong perturbations nearly the entire disc material is removed independent of the investigated
disc-mass distributions. In both cases the mass loss does not depend strongly on the disc-mass distribution. By contrast,
encounters of intermediate strength are most sensitive to the disc-mass distribution. For the case shown in
Fig.~\ref{fig:angularloss}$b$ we find maximum differences of up to $ 35 \, \%$ for $M_2 / M_1 = 3$.  

In case of high-mass ratios $M_2 / M_1 > 20$ and certain non-penetrating periastron distances, which restrict the gravitational
star-disc interactions to the outer disc parts, differences in mass loss of even up to $40 \, \%$ can be inferred for the
different initial disc-mass distributions (see online appendix A).


\subsubsection{Relative angular momentum loss}
\label{subsec:Angular momentum loss}

\begin{figure}[thbp]
  \centering
\includegraphics[width=0.5\textwidth]{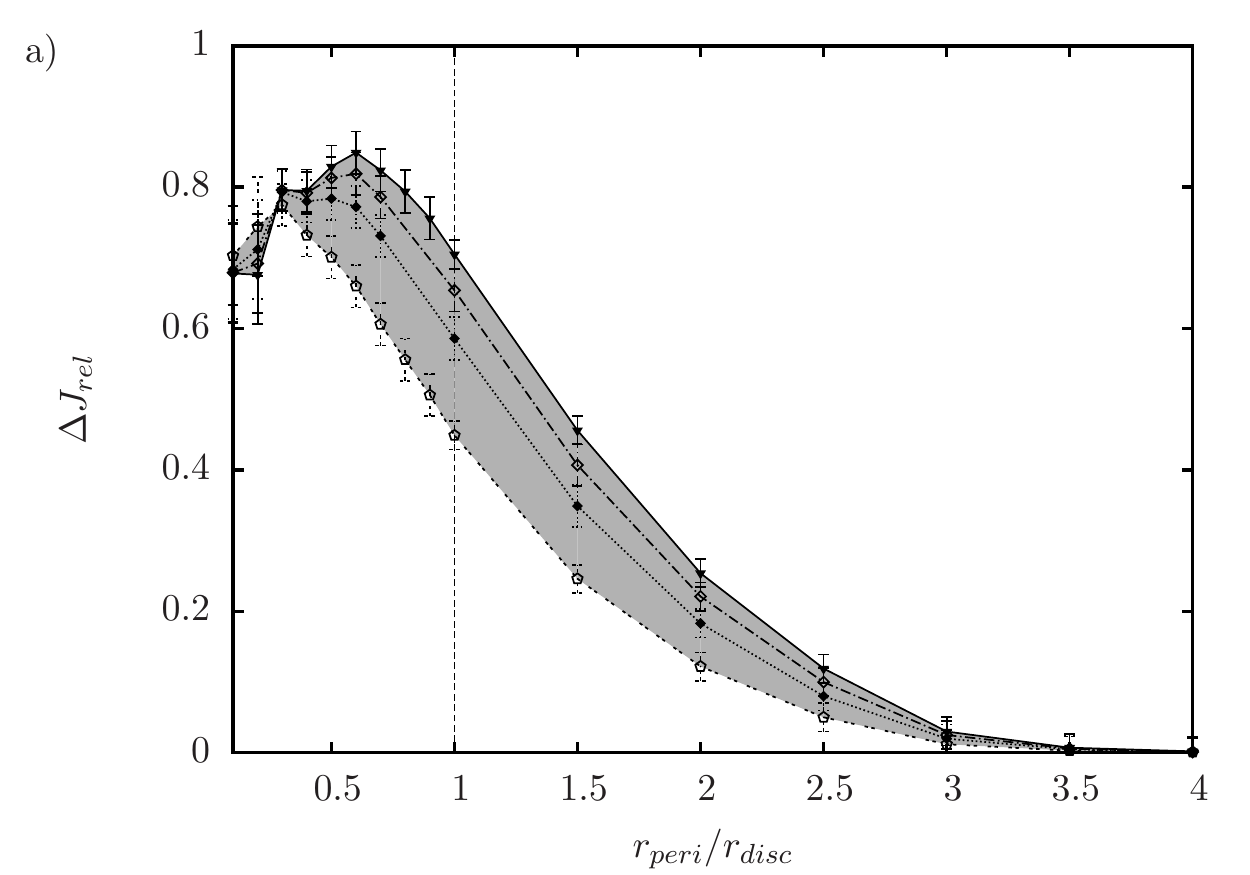}
\includegraphics[width=0.5\textwidth]{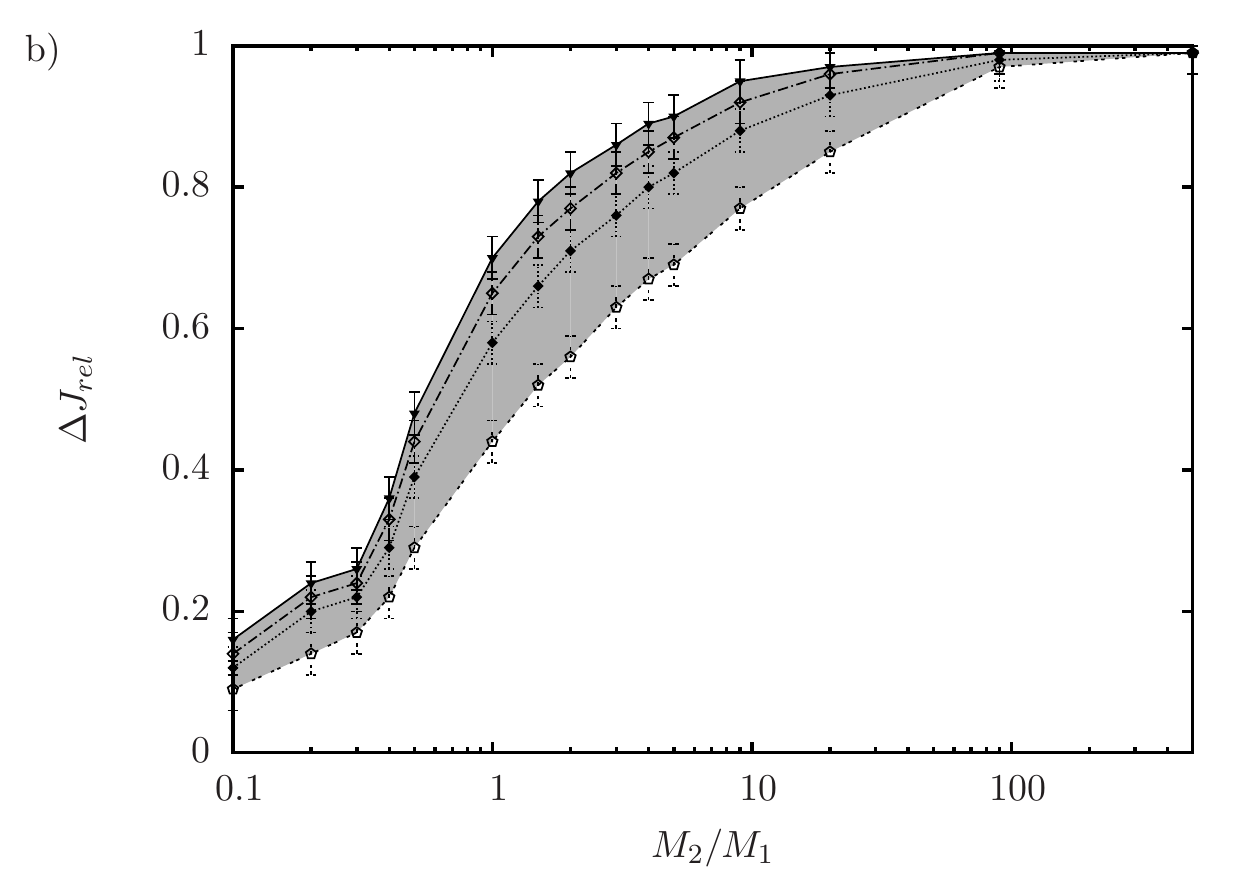}
\caption{The relative angular momentum loss of a $p = 0$ (solid line), $p = 1/2$ (dashed-dotted line), $p = 1$ (dotted line) and
  $p = 7/4$ (double-dotted line) disc-mass distribution including all particles bound tighter to the central star than to the
  perturber and excluding unbound and accreted particles. The data is plotted for $a)$ different periastron distances and an
  equal-mass perturber and for $b)$ different perturber mass ratios and $r_{\mathrm{peri}} / r_{\mathrm{disc}} = 1$. The
  vertical dashed line indicates the initial outer disc radius.}
\label{fig:massloss}
\end{figure}

The different initial disc-mass distributions do not only influence the disc-mass loss but also the angular momentum loss.
Fig.~\ref{fig:massloss}$a$ shows the relative angular momentum loss $\Delta J_{\mathrm{rel}} = ( J_{t=0} - J_{t_{end}}) / J_{t=0}$ as a function
of the encounter distance for equal-mass encounters and the four different disc-mass distributions used in this work. As expected from previous
results \citep[e.g.][]{2007A&A...462..193P} the general trends in relative angular momentum loss are quite similar to that of the
mass loss (compare Fig.~\ref{fig:angularloss} and Fig.~\ref{fig:massloss}) with angular momentum losses being slightly higher than
the disc-mass losses. 
 
  While the mass losses for encounters of intermediate strength with $r_{\mathrm{peri}} / r_{\mathrm{disc}} = 0.9 $ and  $M_2 /
  M_1 = 1$ are $64 \, \%$ for the constant mass distribution compared to $33 \, \%$ for the 7/4-mass
  distribution as shown before, the corresponding angular
  momentum losses are $75 \, \%$  and $50 \, \%$ , respectively. 
  Material migrating inwards or becoming unbound due to an encounter leads to a relative angular momentum loss, while part of
  the disc-mass which is pushed outwards the initial disc-radius but remains bound to the central star increases the total angular
  momentum of the disc. 
  In total, the dependence on the initial disc-mass distributions is less pronounced for the relative angular momentum loss than for the
  disc-mass loss. 

However, although the differences are lower in maximum the influence of different initial disc-mass distributions on the angular
momentum loss covers a significantly large parameter range (see also online appendix B).
\\

  Since mass and angular momentum losses are generally influenced by perturbations of the outer disc parts an initially flat
  particle distribution with high resolution of the outer disc regions results in an increased accuracy of the
  disc losses. Nevertheless, disc losses obtained with initially steep particle distributions ($p = 1$) found in previous studies
  \citep{2006ApJ...642.1140O, 2007A&A...462..193P} are completely reproduced within the error range in the present study.


\subsubsection{Adapting a fit formula dependent on the initial disc-mass distribution}
\label{sec:Adapting of a fit formula}

The numerical results for the mass and angular momentum loss in this study cover a wide parameter range but, however, present only
a discrete classification of the relative losses for the different initial disc-mass distributions.
Analytical approaches are a possible option to avoid this disadvantage but are only valid for a very limited parameter space of
distant encounters  \citep{1994ApJ...424..292O, 2010ApJ...725..353D}  . To obtain a general estimate of the effect of arbitrary initial disc-mass distributions
on the encounter-induced mass and angular momentum loss of protoplanetary discs we present a fit formula that is valid for {\em any
  initial disc-mass distribution}, given that it can be expressed by a power law of the form $\Sigma (r) \propto r^{-p}$. For this
purpose we extended the fit function for the relative mass loss by \citet{2006ApJ...642.1140O} (see Eq.4 therein) and for the
relative angular momentum loss by \citet{2007A&A...462..193P} (see Eq.1 therein), which are valid for a
$r^{-1}$-distribution, towards arbitrary surface density distribution indices $0 \leq p \leq 7/4$: 

\begin{eqnarray}
\Delta m_{\mathrm{rel}} &=&  \left ( \frac{M_2}{M_{2}+0.5M_1} \right )^{1.2} \ln \left[ (3-\frac{p}{4}) \cdot \left ( r_p \right )^{0.1} \right]\\
&\times&\exp \left \{ - \sqrt{ \frac{M_1}{M_2 + 0.5M_1} } \left [ \left ( r_{p} \right )^{2-p/2} -0.5 \right ] \left (
    p+|\frac{1-p}{2}| \right ) \right \} \nonumber 
\label{fitfunc}
\end{eqnarray}

\begin{eqnarray}
  \label{eq:angular_fit}
  \Delta J_{\mathrm{rel}} &=& 1.02 \left [ \frac{M_2}{M_{tot}} +\left (  \frac{1-p}{5}-\frac{\exp(p)}{100}  \right )   \right ]^{0.5 \cdot r_p}
  \nonumber \\
  &\times&\exp \left ( - \sqrt{\frac{M_1(r_p-0.7 \, r_p^{0.5})^3}{M_2}} \right ) \mathrm{.}
\end{eqnarray}
where $r_p=r_{peri}/r_{disc}$ is the relative periastron distance.

For most of the parameter space the adopted functions fit the data well inside the error range and extend Ostriker's analytical
function for the angular momentum loss considerably. 
Larger deviations of the fit functions from the simulated losses occur only for high
encounter mass ratios of $M_2/M_1 > 20$, which was also the case for the established fit functions by \citet{2007A&A...462..193P}
and \citet{2006ApJ...642.1140O}. 
In case of disc-penetrating orbits and low-mass ratios the disc losses can be affected by Lindblad and Corotation resonances that
are located in the inner disc regions and cause moderate deviations of the fit function from the expected disc losses.

The extended fit functions provide a significant improvement of previous analytical and numerical results.
They cover a huge parameter space of the most reasonable initial disc-mass distributions and the most relevant orbital parameter
ranges expected for interactions in star clusters of any age.


\section{Discussion and conclusion}
\label{ch:Conclusion}

Most stars are believed to form in clusters \citep{2003ARA&A..41...57L} and    probably   undergo at least one encounter closer than $1000 \AU$
during the lifetime of their disc ($\sim 10^6$yrs) \citep{2001MNRAS.325..449S}. So encounters are likely to have a significant
effect on the disc structure. The scope of this study was to determine which role the initial mass distribution in the disc plays in this
context. A full parameter study of star-disc encounters, as expected to occur in young dense star clusters like the Orion Nebula
Cluster, has been performed with different disc-mass distributions of the form $\Sigma (r) \propto r^{-p}$, $p \in [ 0, 1/2, 1,7/4 ]$, that cover the whole range of observed mass distributions in discs. The main results can be summarized as follows: 

\begin{enumerate}
\item The relative disc-mass loss among the different initial density distributions 
  differs by up to $40 \, \%$
   for the same type of encounter.  The largest differences are associated with strong perturbations of the outer disc edge, i.e. a grazing encounter in case
  of equal stellar masses.
  
\item Although higher amounts of relative angular momentum than disc-mass are lost due to a star-disc encounter, the dependence on
  the initial disc-mass distribution is less pronounced.
  Nevertheless, half of the parameter range where angular momentum loss occurs shows differences among the investigated
  distributions of more than $15 \, \%$.

\item The disc-mass and angular momentum losses due to a  parabolic  encounter can be fitted by a function depending on the perturber mass ratio, the relative periastron distance and the index of the initial disc-mass distribution. 

\item A steepening of the surface density slope is a general effect of an encounter for any initial disc-mass distribution and can
  result in distribution indices of $p > 2$ even in case of initially flat distributions. 

\end{enumerate}

In short, the intuitive result that the flatter the mass distribution the stronger the change of the disc-mass and angular momentum due to encounters has been
quantified in this study. This infers that the importance of encounters in young stellar clusters and their potential to trigger
encounter-induced losses might be reconsidered depending on the dominant initial disc-mass distribution. 

Another consequence of such an encounter is a change of the surface density in the disc  on short time scales .
In this context, the potentially most significant result of this study is that close encounters can provoke density profiles 
steeper than $\Sigma (r) \propto r^{-2}$ independent of the initial disc-mass distributions. 
 As such density profiles are claimed to be the prerequisite to form a planetary system similar to the solar system,
  encounters in the early history of the solar system could have provided these conditions \citep{2010ARA&A..48...47A}.  
%
%
\citet{2007ApJ...671..878D} analytically inferred an initial surface density profile of the solar nebula protoplanetary disc of
roughly $\Sigma (r) \propto r^{-2.2}$ to form the present planets of the solar system. As a mechanism for the formation of such a steep disc-mass density profile he
considered photoevaporation by an external massive star \citep[see also][]{2010ApJ...722.1115M}. \citet{2004ApJ...612.1147K} found
similar results for extrasolar planetary systems suggesting large disc-mass distribution indices of $p = 2.0 \pm
0.5$. 

Our results emphasize that these steep density profiles do not have to exist ab initio or be formed by photoevaporation processes
but that even discs of initially constant distributed material can fulfill the requirements for the formation of a solar system
type planetary system in the inner disc regions after a close encounter.
Additional evidence for such an encounter in case of the early solar system was given by distant
solar system objects on highly eccentric orbits, like the transneptunian object Sedna, and also the sharp outer edge of the solar
system at $\sim 50 \AU$ from the Sun \citep{1999BAAS...31.1095I, 2001ApJ...549L.241A, 2004AJ....128.2564M,
  2004Natur.432..598K}. 

On the other extreme, encounters can not only lead to profile steepening but as well profile flattening. Even unexpected surface distribution
profiles of $p~<~0$ as observed by \citet{2009ApJ...701..260I} can be explained by the influence of a perturbing
star. Non-penetrating encounters of a star-disc system with initially flat distributed disc material easily lead to such density
profiles of the inner disc regions. However, 
photoevaporation as another significant external impact on protoplanetary discs would have to be considered as
an additional driver in these considerations.

Consequently, if all young stars would start out with the same disc density structure, the influence of the cluster environment by means of
encounters in the very early and dense phases of cluster evolution could account for the observed multitude of disc-mass density
profiles.

\bibliographystyle{aa} 
\bibliography{References} 

\end{document}